\documentclass[10pt,letterpaper]{article}
\usepackage{cogsci}
\usepackage{pslatex}
\usepackage{apacite}
\usepackage{float}
\usepackage{graphicx} 
\usepackage{url}
\usepackage{adjustbox}
\usepackage{newtxtext}
\usepackage{amsmath,amssymb,amsthm}
\usepackage{newtxmath} % must come after amsXXX
\usepackage{fancyhdr}
\usepackage{palatino}
\usepackage{lipsum}
\usepackage{algorithm2e}
\usepackage{newtxtext}
\usepackage{appendix}
\usepackage{pdfpages}
\usepackage{listings}
\usepackage{subcaption}
\usepackage{natbib}
\usepackage{fancyhdr}
 
\title{CogSimulator: A Model for Simulating User Cognition \& Behavior with Minimal Data for Tailored Cognitive Enhancement}

\author{
{\bf Weizhen Bian$^{1,4}$}, 
{\bf Yubo Zhou$^{2}$},
{\bf Yuanhang Luo$^{3}$},
{\bf Ming Mo$^{5}$}, 
{\bf Siyan Liu$^{4}$},\\
{\bf YiKai Gong$^{2}$},
{\bf Renjie Wan$^{2}$},
{\bf Ziyuan Luo$^{2}$}, 
{\bf Aobo Wang$^{4\dag}$}\thanks{$\dag$ Corresponding author. Email: aobo.wang@nus.edu.sg}\\
{\bf $^1$ Department of Computer Science, The Hong Kong University of Science and Technology} \\
{\bf $^2$ Department of Computer Science, Hong Kong Baptist University}\\ 
{\bf $^3$ Department of Applied Mathematics, The Hong Kong Polytechnic University} \\
{\bf $^4$ Department of Systems Science, National University of Singapore} \\
{\bf $^5$ Department of Methodology, The London School of Economics and Political Science}
}

\begin{document}

\maketitle

\begin{abstract}

The interplay between cognition and gaming, notably through educational games enhancing cognitive skills, has garnered significant attention in recent years. This research introduces the CogSimulator, a novel algorithm for simulating user cognition in small-group settings with minimal data, as the educational game Wordle exemplifies. The CogSimulator employs Wasserstein-1 distance and coordinates search optimization for hyperparameter tuning, enabling precise few-shot predictions in new game scenarios. Comparative experiments with the Wordle dataset illustrate that our model surpasses most conventional machine learning models in mean Wasserstein-1 distance, mean squared error, and mean accuracy, showcasing its efficacy in cognitive enhancement through tailored game design.\\
\textbf{Keywords:} 
Artificial Intelligence, Education, Group Behaviour, Skill acquisition and learning, Computational Modeling.

\end{abstract}

\section{Introduction}
In recent years, the application of artificial intelligence has expanded across various fields such as music~\citep{bian2023momusic}, gaming~\citep{yin2023ai}, and healthcare~\citep{shaheen2021applications}, demonstrating its broad impact and potential. Concurrently, the relationship between cognition and games has become a hot topic, particularly in research focused on using games to assess cognitive abilities and explore whether games can enhance cognitive skills~\citep{boot2015video}. At the same time, the relationship between cognitive level and education is also the focus of attention, where cognitive ability is considered a crucial predictor of education and socioeconomic achievement, especially regarding strategies and methods in student learning, individual differences, etc~\citep{yen2004does, van2019use}.\\

In terms of using games to help measure cognitive ability, games can test different aspects of cognitive function, including aspects such as memory and attention, which can assist in diagnosing disease~\citep{wiley2021making}. At the same time, games are thought to be more effective in assessing cognitive abilities, even in improving fairness and user experience~\citep{leutner2023game}. Therefore, games are closely linked to cognitive ability, and in the framework of the importance of cognitive ability in education, the better use of games to contribute to the development of cognitive ability is a significant agenda. However, given the heterogeneous attributes of participants, a fixed game algorithm may not enhance the cognitive abilities of all individuals consistently~\citep{manzano2021between}. Conventional approaches, encompassing behavior trees~\citep{hecker2011my} and data-driven methodologies~\citep{kim2020data}, attempt to tailor game modes to diverse cognitive profiles of users yet necessitate substantial data acquisition. Technologies related to artificial intelligence, including Transfer Learning and Convolutional Neural Networks (CNN), are also challenged by over-fitting issues, consequently diminishing the precision in predicting user cognitive patterns~\citep{zhao2017research}.\\

This work developed the CogSimulator model, which intends to capture and simulate the user's cognitive level based on a small amount of data. Further, using less data to simulate user groups, the game can implement a more targeted educational game for specific users. This paper uses the game ``Wordle", a word game, to demonstrate how to evaluate and improve cognitive ability with a limited number of users, making the game a more acceptable form for teenagers in learning~\citep{amory2010learning}. It is important to highlight that due to the CogSimulator's efficient use of data, encompassing common word frequency metrics pervasive across numerous word-based games, it is well-suited for application to newly developed games or those aimed at niche user groups. By simulating individual player behaviors, the model offers game designers a valuable tool for tailoring game difficulty to match specific cognitive profiles.\\

\section{Related Work}
Cognition includes the mental processes of acquiring knowledge and understanding through thought, experience, and the senses and plays a fundamental role in human development and interaction with the environment~\citep{sommerville2020social}. This area, crucial for critical thinking, problem-solving skills, and the effective processing and interpretation of information, has attracted increasing attention in contemporary research, particularly at the intersection of cognition and play~\citep{majuri2018gamification,sailer2020gamification}. Human research on educational games dates back to 1981 when Malone investigated in his seminal paper how to use the captivating effects of computer games to make learning fun and interesting ~\citep{malone1981toward}. In the mid-1980s, research examined the link between video game play and cognitive performance, and this correlation was gradually confirmed~\citep{dale2020new}. However, even using AI-based game-based educational technologies, such as AI applications for learning new skills or knowledge, may only improve learners' cognition if they motivate long-term use~\citep{laine2020designing}. Therefore, the challenge is producing a game design that suits the needs and preferences of the players to ensure the gamers' cognitive enhancement.\\

Many academic studies have shown that positive motivation is crucial in enhancing human engagement, which in turn helps promote cognitive improvement~\citep{teixeira2012emotion,ryan2000self,dweck2006mindset}. In this context, it can be broken down participation motivation into two categories: intrinsic participation motivation (stemming from the game's intrinsic design elements) and extrinsic participation motivation (stemming from the game's reward and punishment system)~\citep{laine2020designing}. Although the persistence of intrinsically motivated engagement is remarkable, intrinsic motivation depends on various factors, including player type, specific educational needs, and personal interests~\citep{manzano2021between}. Therefore, if the game mode cannot change with different individuals, the degree of cognitive enhancement may vary widely between individuals. This variability poses a significant challenge to developing fixed game mechanics that can universally meet the diverse needs of all players~\citep{manzano2021between}. \\

Traditional approaches to creating adaptive and responsive gaming environments, designed to cater to players' individual needs, have primarily relied on rule-based systems such as finite state machines and behavior trees~\citep{park2023generative}. These systems provide an easy way to build simple agents that provide different feedback based on user actions~\citep{hecker2011my}. However, while this approach effectively creates a baseline interactive experience, it cannot dynamically adapt to individual players' subtle and changing preferences or capabilities. An alternative solution lies in data-driven approaches~\citep{kim2020data}. This approach entails collecting extensive gameplay data, such as average completion times, to assess game difficulty and subsequently recommend games of varying difficulty levels to different players. However, this approach relies on substantial data accumulation for accurate predictions, making it challenging to apply effectively in scenarios requiring rapid adaptation to new tasks. As for artificial intelligence technology, most of the gamification research that appeared in the past decade from 2010 to 2020 failed to provide a structured overview of game elements such as NPCs working with artificial intelligence~\citep{funk2020gamification}. Until 2023, Generative Agents realized the simulation of human behavior based on large language models (LLMs), even in zero-shot scenarios~\citep{park2023generative}. However, the need for extensive resources poses a significant obstacle, especially for small educational games in their embryonic stages~\citep{jozefowicz2016exploring}. For these emerging games, the high threshold of data and computing requirements makes it challenging to fully exploit the potential of AI-driven interactive elements. In contrast, models such as CNN may cause overfitting problems due to insufficient data volume. Therefore, while advances in artificial intelligence technology offer promising avenues for enhancing the realism and engagement of game environments, their applicability still needs to be improved in resource-constrained settings.\\

This work will elucidate and test the model using Wordle as an illustrative case. Wordle is a word-guessing game that epitomizes its players' diverse needs and individual characteristics, reflecting the unique responses and strategies each person brings to the game. The goal of Wordle is to guess a five-letter word within six attempts. After each  guess, the color gives feedbacks as: Green (the correct letter in the right spot), Yellow (The correct letter but in the wrong position) or Grey (the letter outside the word)~\citep{Wordle}. Unaided players guess words mainly through word recall, largely limited by their vocabulary. Consequently, the CogSimulator was developed to emulate user cognition utilizing a limited dataset of game records, thereby facilitating the design of game difficulties optimized for cognitive enhancement.\\

\section{CogSimulator Model}

\begin{figure*}[h]
\centering
\includegraphics[width=0.85\textwidth]{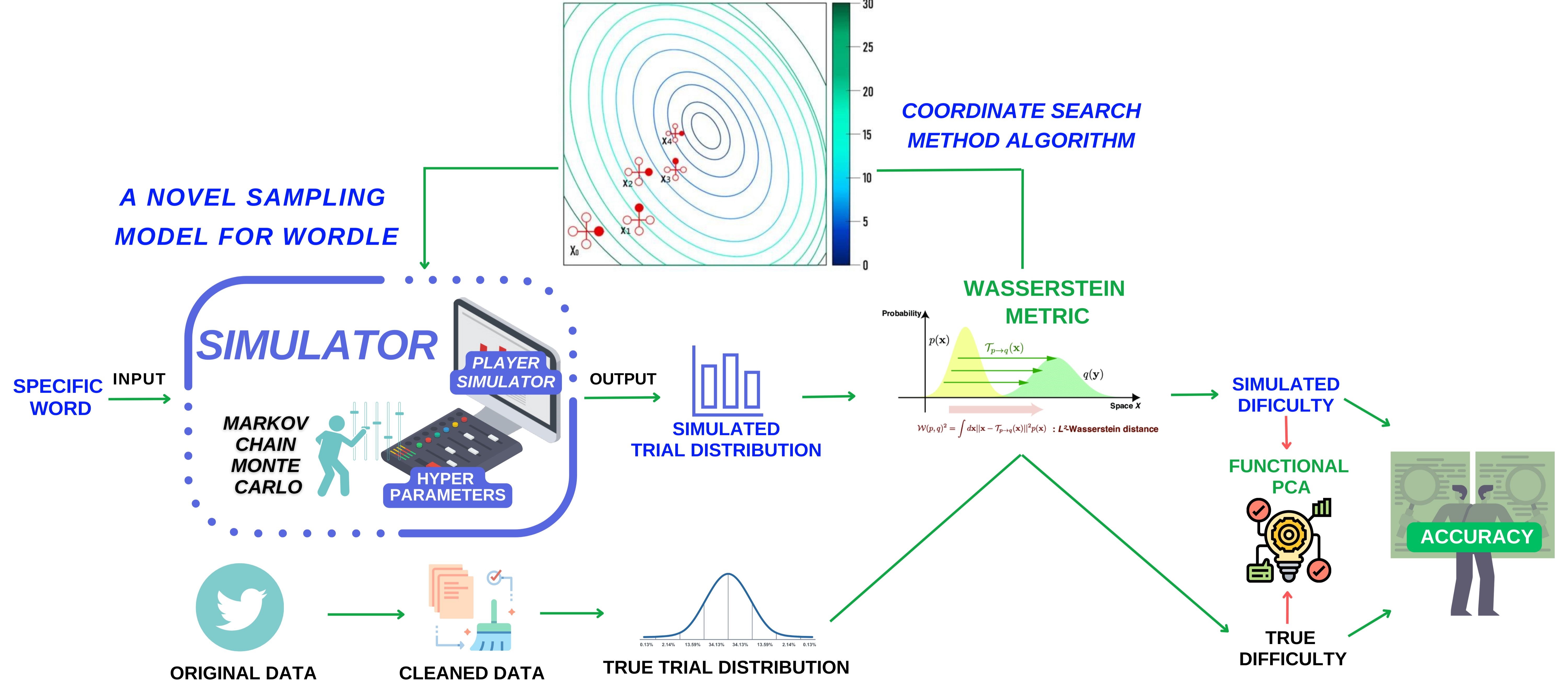}
\caption{An overview of the CogSimulator model. It starts by aggregating Wordle data from Twitter and using it to form a true trial distribution for training. The simulator, via Markov Chain Monte Carlo, adjusts hyperparameters to match this distribution. Parameter optimization is performed with a Coordinate Search Method Algorithm, and model accuracy is ensured using the Wasserstein-1 metric to compare simulated outputs to actual trials. Functional PCA relates simulated and true word difficulties, enabling precise predictions for new word trials based on players' cognitive patterns.}
\label{overview}
\end{figure*}

\subsection{Model Overview}

We analyze that traditional machine learning models often face severe overfitting due to insufficient sample sizes, limiting their effectiveness. In contrast, our model excels in explaining word difficulty and result distribution, offering a more robust solution as detailed in Table \ref{tab:my_label}. Attributes such as word frequency, prevalent across various word-based games, provide a solid foundation for applying our model to different gaming scenarios. To address these challenges, we have developed a novel sampling simulator that better reflects the gameplay dynamics of the broader population, as illustrated in Figure \ref{overview}.\\

The Cogsimulator operates through a process analogous to Markov Chain Monte Carlo (MCMC) as described by Geyer~\citep{geyer2011introduction}, where parameters progressively converge to a steady state distribution and achieve detailed balance. The model incorporates hyperparameters that capture players' cognitive processes at each stage of their guessing attempts, alongside the stochastic variations observed in each trial. Unlike traditional optimization methods that require derivatives of the cost function, our approach uses Coordinate Search Optimization~\citep{frandi2014coordinate}, which optimizes parameters coordinate-wise within the hyperparameter space in each iteration, allowing the model to align with the actual trial distributions observed in data eventually. To measure the deviation between our training set and the outputs generated by the simulation, we employ the Wasserstein-1 distance~\citep{liao2022fast}, whose convex properties aid in steering the algorithm towards the global optimum. This setup ensures our model accurately estimates word difficulty and adeptly classifies the distribution of players' guessing attempts. Once optimized, the sampling simulator effectively replicates samples that reflect the average performance of the simulated player population.\\

Further refining our approach, we estimate the difficulty of newly introduced words by comparing their Wasserstein distance with that of the most straightforward word identified in the training dataset. This comparison uses the generated Wasserstein Metric to provide a relative measure of word complexity. Consequently, when a new word is introduced to the model, it predicts not only the difficulty of the word but also the expected number of guesses by players. This prediction is based on the collective cognitive profile of the player group derived from the training data, thereby aiding in selecting words that are optimally challenging yet engaging for players.

\subsection{Simulator}
As the core component of CogSimulator, the simulator uses a coordinate search method algorithm to tune hyperparameters to match human performance in player simulators automatically. The simulator takes a single word as input and can generate a distribution of the number of times a user guessed the word, containing the percentage of attempts for each of the seven trial types. Other factors will be incorporated into the neural network alongside word guessing frequency distribution. This integration will occur through a process of 5-fold cross-validation to ensure a better fit for various real-world factors, thereby yielding the most accurate predictions. To determine whether the player selected a specific word in a step, given a qualified dictionary that allows him to choose, algorithms 1 and 2 output the words sampled in a round.\\

\RestyleAlgo{ruled}
\begin{algorithm}[hbt!]
\caption{Choose A Word algorithm $\mathcal{CW}$}\label{alg:CW}
\KwData{Dictionary $\mathcal{D}\ni \{A_i\}^N$ with word frequency $\{p(A_i)\}^N$}
\textbf{Given hyper-parameters: $K, T$} \;
\textbf{Initialise:} Probability list of length $N$: $\operatorname{PL}=[0,0,\ldots,0] $\;
    \While{$i \leq N$}{
  \textbf{update} $\operatorname{PL}[i]\leftarrow \max(0,p(A_i)\times T)$ \;
  % Correction: Removed the \longleftarrow and replaced \max{...} with \max(...)
  % Also, ensure the update statements are ended with \;
  Delete words from $\mathcal{D}$ according to Wordle clue rule (colored clue)\;
  \textbf{update} $\operatorname{PL}[i]\leftarrow \operatorname{normalise}(\operatorname{PL})$ \;
  % Correction: Removed the \r before \operatorname{normalise}
  \textbf{update} $\operatorname{Chosen word}\leftarrow$ randomly choose a word from $\mathcal{D}$ with probability $\operatorname{PL}$\;
  % Correction: Ensured consistency in the use of \leftarrow
 }
 \textbf{Return} $\operatorname{Chosen word}$
\end{algorithm}

\RestyleAlgo{ruled}
\begin{algorithm}[hbt!]
\caption{Single word Trial Simulation Algorithm}\label{alg:ALgo}
\KwData{Any target word $A^*$, dictionary $\mathcal{D}\ni \{A_i\}^N$ with word frequency $\{p(A_i)\}^N$}
\textbf{Given hyper-parameters: $K, T$} \;
\textbf{Initialise:} Count of steps: $\operatorname{cnt}=1 $\;
    \textbf{Set:}$\operatorname{Chosen word}=\mathcal{CW}(\mathcal{D}^*,\{p(A_i)\}^N,K,T$)\;
    
    \While{$\operatorname{Chosen word}\neq A$}{
  \textbf { update } $\mathcal{D}^*\longleftarrow$ Delete words from $\mathcal{D}$ according to Wordle clue rule (colored clue)\;
  
  \textbf { update } $\operatorname{Chosen word}=\mathcal{C}(\mathcal{D},\{p(A_i)\}^N,K,T)$ \;
  
   $\operatorname{cnt}=\operatorname{cnt}+1$ \;
 }
 \textbf{Return} $\operatorname{cnt}$
\end{algorithm}

In simulating Wordle player choices, the \textit{CogSimulator} posits that players are more inclined to guess words they encounter more frequently. However, the human capacity to recall words is finite, and the probability of recalling a specific word is not strictly proportional to its word frequency. Thus, we introduce two parameters to adjust for this. The first parameter, \( K \), represents the cognitive limit, or the maximum number of words a person can typically remember at one time. The simulator only considers up to \( K \) most frequent words as viable options for player guesses. The second parameter, \( T \), represents a scaling factor for the frequency of the most common word, providing a baseline for comparison. To calculate the selection probability for each word, we multiply its frequency by \( T \) and then normalize by dividing over the sum of the scaled frequencies for all \( K \) words considered. This method ensures that while the likelihood of selecting highly frequent words is amplified, less common words maintain a non-zero probability of selection proportional to their relative frequency. This nuanced approach balances the natural human tendency to favor familiar words and the game's challenge to recall less frequent words.

On each trial, the simulator randomly samples words based on their probability of selection~\citep{el2021optimal}. After 1000 samples, a trial distribution of words is generated. At the same time, a coordinate search algorithm will be applied to automatically adjust hyperparameters to improve the fitting accuracy of the simulator. Finally, the simulator can generate seven probability distributions of word guesses containing the percentages for seven categories (1, 2, 3, 4, 5, 6, X). Finally, the model can predict the difficulty of a word for a specific cognitive user through distribution and Wasserstein Metric.

\subsection{Wasserstein Metric}
To test the difficulty of a word for a particular cognitive group, we first obtain the distribution of word guessing times for this cognitive group's best record in past games. Then, by comparing the distance between the distribution of new input words and the distribution of the best record, the difficulty of the word for this cognitive group is judged. For this purpose, we propose using the Wasserstein-1 distance to evaluate the discrepancy between two trial distributions. Let $p, q$ be two probability distributions on compact spaces. Denote $\Pi(p, q)$ as the set of all distributions $\pi(\omega, \omega')$ on $\mathcal{X} \times \mathcal{X}'$ such that the marginals are $p(x)$ and $q(y)$ respectively. Then the Wasserstein-1 distance between $p$ and $q$ is
\begin{equation}
\begin{aligned}
\label{W1}
W_1(p, q)&=\inf _{\pi \in \Pi(p, q)}\int_{\mathcal{X} \times \mathcal{X}}\|\omega-\omega'\| d\pi(\omega, \omega')\\&=\inf _{\pi \in \Pi(p, q)} \underset{(\omega, \omega') \sim \pi}{\mathbb{E}}\left[\|\omega-\omega'\|\right].
\end{aligned}
\end{equation}

\noindent when \ref{W1} applies to discrete sample spaces, let us assume $\mathcal{X}=\left\{\omega_i\right\}_{i=1}^m$ and $\mathcal{X}'=\left\{\omega_i^{\prime}\right\}_{i=1}^{m'}$. $p$ and $q$ are trial distributions of words $A$ and $A*$, respectively. The distance between $p$ and $q$ can be obtained by solving the following linear programming problem

\begin{equation}
\begin{aligned}
\label{LP}
    \mathcal{W}(A, A*) &= \inf_{\{\gamma_{i, j}\}, i, j} \left\{ \sum_{i=1}^m \sum_{j=1}^{m'} \gamma_{i, j}|\omega_i-\omega_j^{\prime}|: \right. \\
    &\left. \sum_{i=1}^s \gamma_{i, j}=q_j, \sum_{j=1}^{s^{\prime}} \gamma_{i, j}=p_i, \gamma_{i, j} \geq 0 \right\}.
\end{aligned}
\end{equation}

 \noindent In this special case where the support of any trial distribution is the same, \emph{i.e.} $m=m'$ with uniform weights, it can be easily shown that Wasserstein-1 distance has a nice closed and compact form $\frac{1}{m} \sum_{i=1}^m|\omega_{\eta(i)}-\omega'_{\vartheta(i)}|$ where $\eta$ is a sorting permutation of $\omega_i$ and $\vartheta$ is a sorting permutation $\omega'_j$. The difficulty of word $A$ is determined by calculating its Wasserstein-1 distance w.r.t trial distribution of the `easiest' in the dataset, which we assume is the word ``train". Therefore, we propose the Wasserstein-1 difficulty measure of the form.
 
 \begin{equation}
 \label{Diff}
     \mathcal{W}^*(A):= \mathcal{W}(p_{\text{``train"}}, p_{A}).
 \end{equation}

\subsection{Coordinate Search Optimization}

To obtain parameters to simulate the cognition of the target population, we used Coordinate Search Optimization. The algorithm is as follows:

\RestyleAlgo{ruled}
\begin{algorithm}[hbt!]
\caption{Coordinate Search algorithm.}\label{alg:ALgo}
\KwData{Ground truth trial distribution $f_{A_i}$,data set word ${A_i}^{355}$, dictionary $\mathcal{D}\ni \{A_i\}^N$ with word frequency $\{p(A_i)\}^N$}
\textbf{Initialise:} Hyper-parameters: $K_0 ,T_0$\;
    
    \While{$j\leq$ \text{maximum iteration (or not converge)}}{

  \textbf { update } $K_j\longleftarrow \text{arg}\min_{K^*} f(K^*,T_{j-1})$ \;
  
    \textbf { update } $T_j\longleftarrow \text{arg}\min_{T^*} f(K_j,T^*)$ \;
 }
 \textbf{Return} $\operatorname{cnt}$
\end{algorithm}

\noindent By limiting the set of search directions to the axes of the input space, the coordinate search/descent technique is an alternative zero-order local approach that addresses the scaling issue seen in traditional local search. The theory is intuitive ~\citep{gurbuzbalaban2020randomness}: random search was created to simultaneously minimize the mean W-1 discrepancy about all of its parameters, which takes the form.
\begin{equation}
    f(T,K)=(\frac{1}{355})\sum_{i=1}^{355}\mathcal{W}(f_{A_i},f^*_{A_i}),
\end{equation}
where $f_{A_1}$ and $f^*_{A_1}$ are trial distribution of target word $A_i$ from ground truth and simulated result, respectively. Note that the distribution $f^*_{A_i}$ is a realization from our Wordle simulator given hyper-parameter $K$ and $T$. A coordinate-wise algorithm reduces this function to one coordinate or weight at a time or, more generally, to a subset of coordinates or weights at a time while holding the other coordinates or weights constant. Despite this limitation, these algorithms are much more versatile than random search (in fact, they may be used to solve even medium-sized machine learning issues efficiently in practice) ~\citep{deshpande2018generative}, even though they require additional steps to establish approximate minima and restrict the number of descent directions that can be discovered. These algorithms also act as predicates for a whole line of higher-order coordinate descent techniques, just like they did with the random search strategy.\\

\section{Experiment}
\subsection{Dataset}
The model employed in this study utilized Wordle results sourced from Wordle Stats over the full year from January 7, 2022, to December 31, 2022. This one year was chosen to capture longitudinal data, reflecting genuine player interactions throughout different seasons and stages of player development, thereby minimizing pre-selection bias and providing a comprehensive basis for understanding group behavior in word-guessing activities. The results include the distribution of the number of trials it took players to succeed, a critical measure of the game's difficulty, and player engagement. These details can be found at \url{https://shorturl.at/adeO6}. Additionally, to construct a robust model, we integrated a dictionary database comprising five-word English terms from the Google Books Ngram Corpus from 1970 to 2019. This corpus provided the foundational data for calculating each word's usage frequency, a key determinant of user selection within our predictive framework. We define word frequency as the relative occurrence of a word in the corpus, which is a direct measure of its commonality and presumed familiarity to players, reflecting the likely cognitive effort required for players to guess the word correctly~\citep{solovyev2019google}.

\subsection{Model Evaluation}

\subsubsection{Wasserstein Metric}

\begin{figure}[htbp]
\centering
\includegraphics[width=0.48\textwidth]{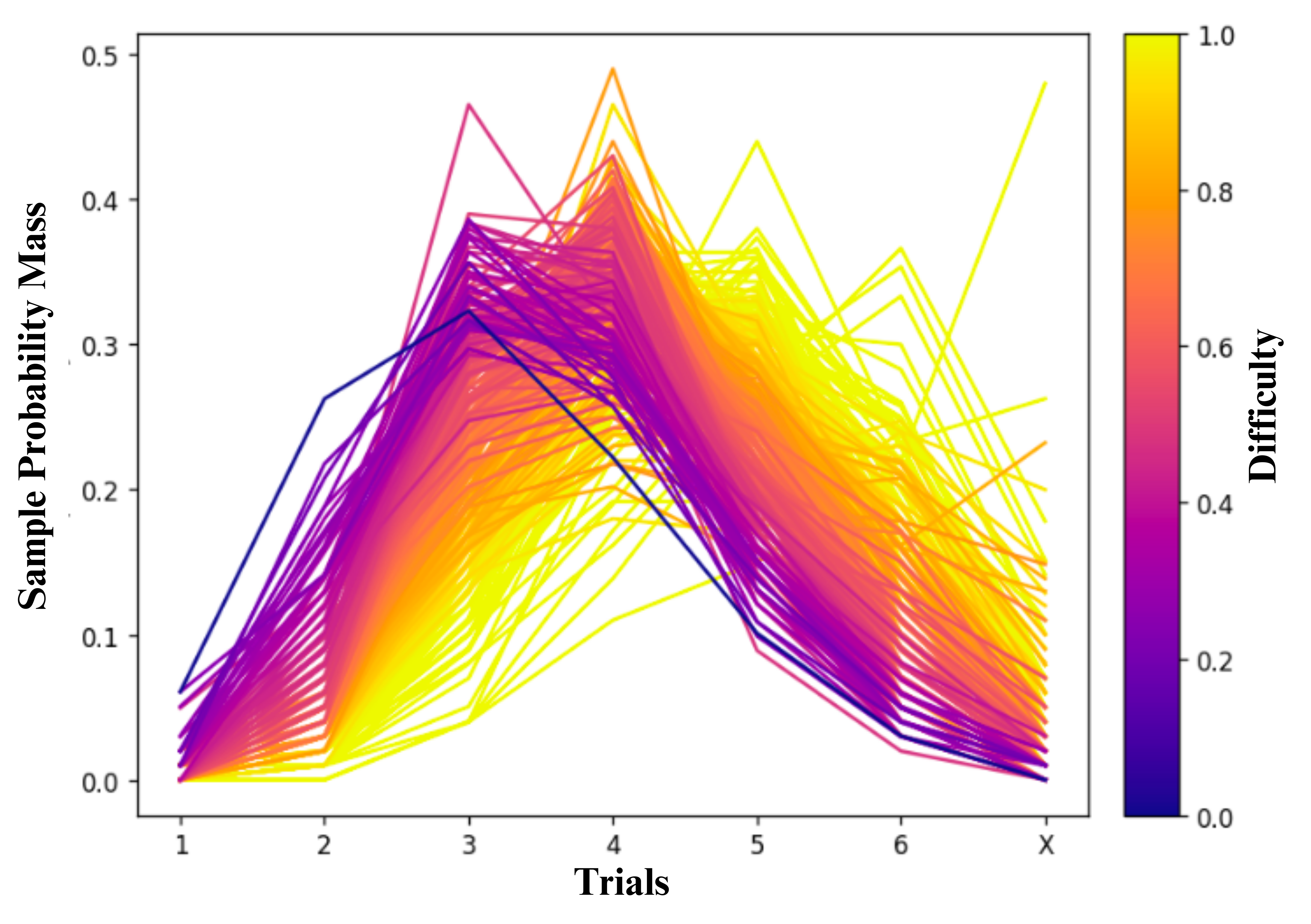}
\caption{Distribution of word difficulties in the Wordle game. The x-axis represents the number of attempts, and the y-axis represents the frequency of each attempt number.}
\label{fig:diff}
\end{figure}

This algorithm uses Wasserstein-1 distance to evaluate the difference between two trial distributions. Here, the attempt distribution refers to the attempts the user requires to complete the task in the Wordle game. This model aims to determine the difficulty of the word A relative to the word A$^*$ in the Wordle game. This is done by calculating the Wasserstein-1 distance of word A relative to the overall attempt distribution. Figure \ref{fig:diff} shows the difficulty distribution of 355 ground truth words. It shows that the proposed metric is sufficiently consistent, representative of the difficulty realized by records, and mediates between other quantifiers. Clearly, `easier' target words exhibit a trial distribution shifted towards the left on the x-axis, and `harder' target words ( yellow colored) have a right-shifted trail distribution.

\subsubsection{Coordinate Search Optimization}

\begin{figure}[hbt!]
    \centering
    \begin{subfigure}[b]{0.73\columnwidth}
        \centering
        \includegraphics[width=\textwidth]{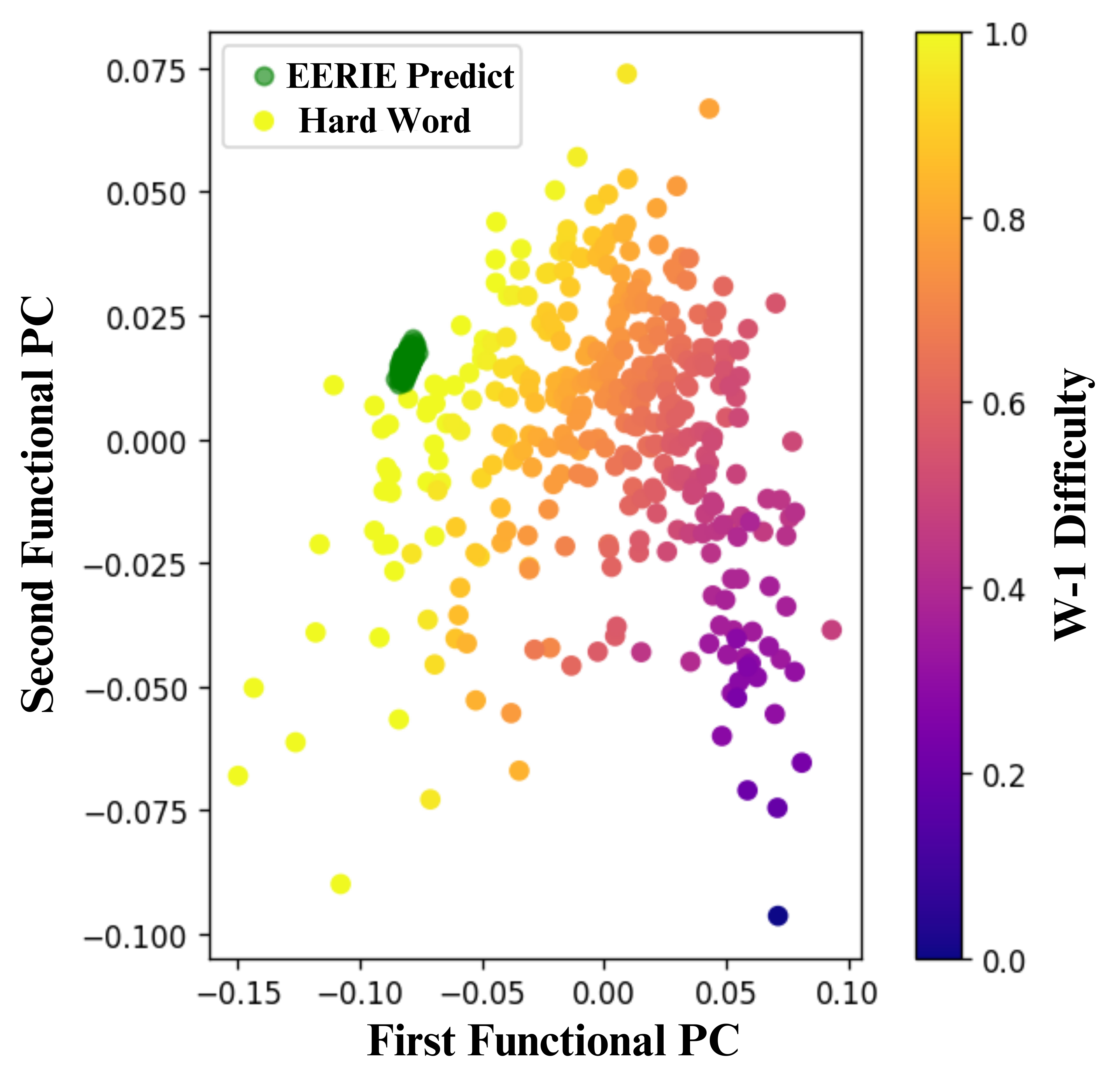}
        \label{fig:predict}
    \end{subfigure}

    \begin{subfigure}[b]{0.73\columnwidth}
        \centering
        \includegraphics[width=\textwidth]{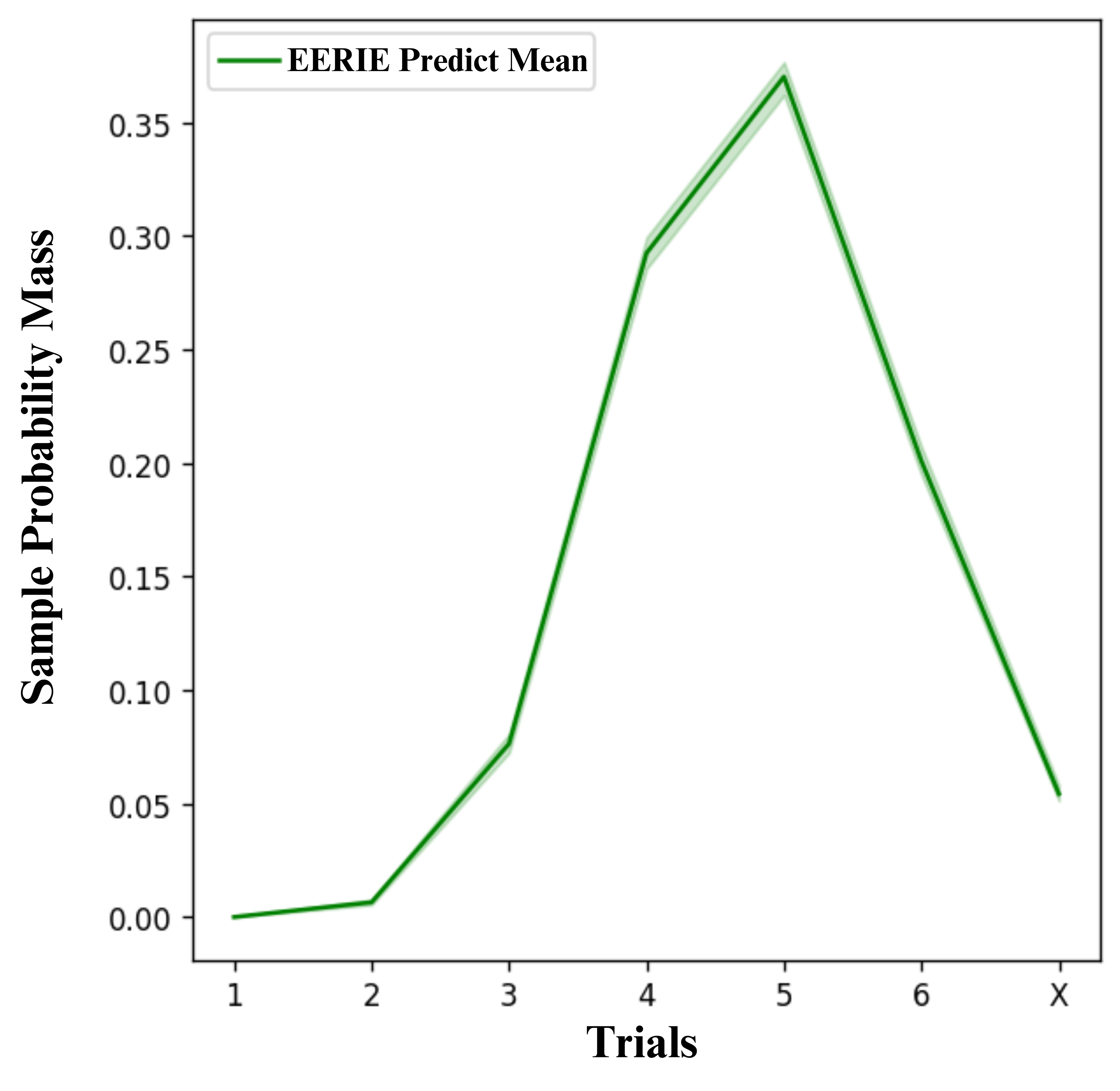}
        \label{fig:predict_confidence}
    \end{subfigure}
     \caption{We obtain 200 simulated trial distributions of the word eerie. Displayed at the top, the visualization method projects these results into a 2-dimensional functional space. At the bottom, the predicted mean is highlighted with a green line, and a $95\%$ empirical confidence interval is constructed around it.}
    \label{fig:pca_fit}
    
\end{figure}

To assess the efficacy of the Coordinate Search Optimization algorithm, it is essential to consider the broader statistical profile rather than just a single winning result. This algorithm's utility lies in its ability to iteratively explore and optimize a multi-dimensional space by adjusting one coordinate at a time. It is particularly suited for problems with a complex objective function or lack an analytical gradient. The evaluation of this optimization technique is predicated on its capacity to train generators that output a discrete target trial distribution accurately. We plan to invoke the same number of generative realizations for a robust evaluation as in fixed-length accurate data batches. This approach ensures a fair comparison between the model's output and the empirical ground truth. Consistency with the ground truth is then assessed by visualizing the distribution of densities using the same projection of functional PCA~\citep{shang2014survey}, as depicted in Figure \ref{fig:pca_fit} (a). Further, we solidify our statistical analysis by constructing an empirical 95\% confidence interval for the optimization outcomes, with the results presented in Figure \ref{fig:pca_fit} (b) indicating that the sampling algorithm is robust, as evidenced by the slight variance observed. The predictions and summaries derived from this approach are systematically tabulated for further scrutiny.\\

\subsubsection{Simulator}

The model is benchmarked with several other machine learning algorithms that comprehensively estimate trial distributions. Note that we are using the attributes to feed as the input; in regression-type benchmarks, the machine learning algorithms output a $7$ dimensional vector filled with floating number loadings, whereas in classification tasks, benchmarks output a category. We train these machine learning algorithms under the canonical parameter settings. Representative methods, namely, linear regression, decision tree regression, random forest regression, and Multiple layer perceptron regressor, are selected for the experiments, and the actual performance of the algorithms is compared by studying the training and validating performance among these algorithms. For the sake of fairness, hyperparameters and settings for each method are set by default.\\

The simulator has proven a successful method for predicting the difficulty and distribution of words in the simulator game. The model achieves an accuracy of 87$\%$, outperforming other machine learning algorithms we have tried. The results demonstrate that the simulator is effective in predicting the distribution of future reports and the difficulty and distribution of specific words in Wordle.\\

\setlength{\tabcolsep}{1.4mm}{
\begin{table}[h]
    \centering
    \caption{For distribution learner, We report mean squared loss for all models. Mean accuracy is obtained from 5-fold cross-validation; note that our model outperforms all benchmarks.}
    \begin{tabular}{ccc}
\hline Model & MSE & Mean Acc \\
\hline CogSimulator            & 0.512 & $\mathbf{87\%}$ \\
Linear Regression        & 1.206 & 19\%\\
MLP Regression           & 1.724 & 62\%\\
Regression Tree          & 0.758 & 60\%\\
RandomForest Regression & 0.620 & 68.45\%  \\
\hline
\end{tabular}
    
    \label{tab:my_label}
\end{table}}

Overall, the simulator significantly improves the prediction of word difficulty and distribution in Wordle. Its superior performance compared to other machine learning algorithms highlights the effectiveness of the Convolutional Networks under the Wasserstein training approach. This work proposes a robust model for predicting the difficulty of educational games tailored to specific cognitive groups, ultimately enhancing overall game playability and engagement. The results of different models are shown in the table \ref{tab:my_label}.\\

\section{Conclusion and Future Work}

This study introduces the CogSimulator, designed to simulate cognitive distributions in contexts with limited sample sizes, exemplified by its application to Wordle. The model competes against machine learning metrics such as Wasserstein-1 distance, mean squared error, and mean accuracy. Leveraging the universal relevance of word frequency attributes, the CogSimulator shows promising generalization across word-based games, suggesting significant potential impacts on cognitive game development for niche user groups. Despite its strengths, the model tends to represent an ``average" player, which may not reflect the diversity in player strategies and cognitive processes, particularly where data distributions are multimodal.\\

Future work will enhance this model by integrating clustering algorithms to detect and model unique player profiles, thereby better capturing the diversity of players. Plans also include integrating player feedback as a dynamic reward mechanism and developing an advanced parameter optimization method for dynamic loss functions to suit simulation tasks better and extend applicability to a broader range of educational games.

\nocite{ChalnickBillman1988a}
\nocite{Feigenbaum1963a}
\nocite{Hill1983a}
\nocite{OhlssonLangley1985a}
% \nocite{Lewis1978a}
\nocite{Matlock2001}
\nocite{NewellSimon1972a}
\nocite{ShragerLangley1990a}

\bibliographystyle{apacite}

\setlength{\bibleftmargin}{.125in}
\setlength{\bibindent}{-\bibleftmargin}

\bibliography{CogSci_Template}

\begin{thebibliography}{}

\bibitem [\protect \citeauthoryear {%
Amory%
}{%
Amory%
}{%
{\protect \APACyear {2010}}%
}]{%
amory2010learning}
\APACinsertmetastar {%
amory2010learning}%
\begin{APACrefauthors}%
Amory, A.%
\end{APACrefauthors}%
\unskip\
\newblock
\APACrefYearMonthDay{2010}{}{}.
\newblock
{\BBOQ}\APACrefatitle {{Learning} to {Play} {Games} or {Playing} {Games} to {Learn}? {A} {Health} {Education} {Case} {Study} with {Soweto} {Teenagers}} {{Learning} to {Play} {Games} or {Playing} {Games} to {Learn}? {A} {Health} {Education} {Case} {Study} with {Soweto} {Teenagers}}.{\BBCQ}
\newblock
\APACjournalVolNumPages{Australasian Journal of Educational Technology}{26}{6}{}.
\PrintBackRefs{\CurrentBib}

\bibitem [\protect \citeauthoryear {%
Bian%
\ \protect \BOthers {.}}{%
Bian%
\ \protect \BOthers {.}}{%
{\protect \APACyear {2023}}%
}]{%
bian2023momusic}
\APACinsertmetastar {%
bian2023momusic}%
\begin{APACrefauthors}%
Bian, W.%
, Song, Y.%
, Gu, N.%
, Chan, T\BPBI Y.%
, Lo, T\BPBI T.%
, Li, T\BPBI S.%
\BDBL {}Trillo, R\BPBI A.%
\end{APACrefauthors}%
\unskip\
\newblock
\APACrefYearMonthDay{2023}{}{}.
\newblock
{\BBOQ}\APACrefatitle {{MoMusic}: A {M}otion-{D}riven {H}uman-{AI} {C}ollaborative {M}usic {C}omposition and {P}erforming {S}ystem} {{MoMusic}: A {M}otion-{D}riven {H}uman-{AI} {C}ollaborative {M}usic {C}omposition and {P}erforming {S}ystem}.{\BBCQ}
\newblock
\BIn{} \APACrefbtitle {Proceedings of the {AAAI} Conference on Artificial Intelligence} {Proceedings of the {AAAI} conference on artificial intelligence}\ (\BVOL~37, \BPGS\ 16057--16062).
\PrintBackRefs{\CurrentBib}

\bibitem [\protect \citeauthoryear {%
Boot%
}{%
Boot%
}{%
{\protect \APACyear {2015}}%
}]{%
boot2015video}
\APACinsertmetastar {%
boot2015video}%
\begin{APACrefauthors}%
Boot, W\BPBI R.%
\end{APACrefauthors}%
\unskip\
\newblock
\APACrefYearMonthDay{2015}{}{}.
\newblock
\APACrefbtitle {{Video} {Games} as {Tools} to {Achieve} {Insight} into {Cognitive} {Processes}} {{Video} {Games} as {Tools} to {Achieve} {Insight} into {Cognitive} {Processes}}\ (\BVOL~6).
\newblock
\APACaddressPublisher{}{Frontiers Media SA}.
\PrintBackRefs{\CurrentBib}

\bibitem [\protect \citeauthoryear {%
Dale%
, Joessel%
, Bavelier%
\BCBL {}\ \BBA {} Green%
}{%
Dale%
\ \protect \BOthers {.}}{%
{\protect \APACyear {2020}}%
}]{%
dale2020new}
\APACinsertmetastar {%
dale2020new}%
\begin{APACrefauthors}%
Dale, G.%
, Joessel, A.%
, Bavelier, D.%
\BCBL {}\ \BBA {} Green, C\BPBI S.%
\end{APACrefauthors}%
\unskip\
\newblock
\APACrefYearMonthDay{2020}{}{}.
\newblock
{\BBOQ}\APACrefatitle {{A} {New} {Look} at the {Cognitive} {Neuroscience} of {Video} {Game} {Play}} {{A} {New} {Look} at the {Cognitive} {Neuroscience} of {Video} {Game} {Play}}.{\BBCQ}
\newblock
\APACjournalVolNumPages{Annals of the New York Academy of Sciences}{1464}{1}{192--203}.
\PrintBackRefs{\CurrentBib}

\bibitem [\protect \citeauthoryear {%
Deshpande%
, Zhang%
\BCBL {}\ \BBA {} Schwing%
}{%
Deshpande%
\ \protect \BOthers {.}}{%
{\protect \APACyear {2018}}%
}]{%
deshpande2018generative}
\APACinsertmetastar {%
deshpande2018generative}%
\begin{APACrefauthors}%
Deshpande, I.%
, Zhang, Z.%
\BCBL {}\ \BBA {} Schwing, A\BPBI G.%
\end{APACrefauthors}%
\unskip\
\newblock
\APACrefYearMonthDay{2018}{}{}.
\newblock
{\BBOQ}\APACrefatitle {{Generative} {Modeling} {Using} the {Sliced} {Wasserstein} {Distance}} {{Generative} {Modeling} {Using} the {Sliced} {Wasserstein} {Distance}}.{\BBCQ}
\newblock
\BIn{} \APACrefbtitle {Proceedings of the IEEE Conference on Computer Vision and Pattern Recognition} {Proceedings of the ieee conference on computer vision and pattern recognition}\ (\BPGS\ 3483--3491).
\PrintBackRefs{\CurrentBib}

\bibitem [\protect \citeauthoryear {%
Dweck%
}{%
Dweck%
}{%
{\protect \APACyear {2006}}%
}]{%
dweck2006mindset}
\APACinsertmetastar {%
dweck2006mindset}%
\begin{APACrefauthors}%
Dweck, C\BPBI S.%
\end{APACrefauthors}%
\unskip\
\newblock
\APACrefYear{2006}.
\newblock
\APACrefbtitle {{Mindset}: {The} {New} {Psychology} of {Success}} {{Mindset}: {The} {New} {Psychology} of {Success}}.
\newblock
\APACaddressPublisher{}{Random House}.
\PrintBackRefs{\CurrentBib}

\bibitem [\protect \citeauthoryear {%
El-Hadidy%
, Alfreedi%
\BCBL {}\ \BBA {} Alzulaibani%
}{%
El-Hadidy%
\ \protect \BOthers {.}}{%
{\protect \APACyear {2021}}%
}]{%
el2021optimal}
\APACinsertmetastar {%
el2021optimal}%
\begin{APACrefauthors}%
El-Hadidy, M\BPBI A\BPBI A.%
, Alfreedi, A\BPBI A.%
\BCBL {}\ \BBA {} Alzulaibani, A\BPBI A.%
\end{APACrefauthors}%
\unskip\
\newblock
\APACrefYearMonthDay{2021}{}{}.
\newblock
{\BBOQ}\APACrefatitle {{Optimal} {Multiplicative} {Generalised} {Coordinated} {Search} {Technique} to {Find} a {D-dimensional} {Random} {Walker}} {{Optimal} {Multiplicative} {Generalised} {Coordinated} {Search} {Technique} to {Find} a {D-dimensional} {Random} {Walker}}.{\BBCQ}
\newblock
\APACjournalVolNumPages{International Journal of Operational Research}{42}{1}{1--33}.
\PrintBackRefs{\CurrentBib}

\bibitem [\protect \citeauthoryear {%
Frandi%
\ \BBA {} Papini%
}{%
Frandi%
\ \BBA {} Papini%
}{%
{\protect \APACyear {2014}}%
}]{%
frandi2014coordinate}
\APACinsertmetastar {%
frandi2014coordinate}%
\begin{APACrefauthors}%
Frandi, E.%
\BCBT {}\ \BBA {} Papini, A.%
\end{APACrefauthors}%
\unskip\
\newblock
\APACrefYearMonthDay{2014}{}{}.
\newblock
{\BBOQ}\APACrefatitle {{Coordinate} {Search} {Algorithms} in {Multilevel} {Optimization}} {{Coordinate} {Search} {Algorithms} in {Multilevel} {Optimization}}.{\BBCQ}
\newblock
\APACjournalVolNumPages{Optimization Methods and Software}{29}{5}{1020--1041}.
\PrintBackRefs{\CurrentBib}

\bibitem [\protect \citeauthoryear {%
Funk%
, Dieber%
, Pichler%
\BCBL {}\ \BBA {} Coeckelbergh%
}{%
Funk%
\ \protect \BOthers {.}}{%
{\protect \APACyear {2020}}%
}]{%
funk2020gamification}
\APACinsertmetastar {%
funk2020gamification}%
\begin{APACrefauthors}%
Funk, M.%
, Dieber, B.%
, Pichler, H.%
\BCBL {}\ \BBA {} Coeckelbergh, M.%
\end{APACrefauthors}%
\unskip\
\newblock
\APACrefYearMonthDay{2020}{}{}.
\newblock
{\BBOQ}\APACrefatitle {{Gamification} of {Trust} in {HRI}?} {{Gamification} of {Trust} in {HRI}?}{\BBCQ}
\newblock
\BIn{} \APACrefbtitle {Culturally Sustainable Social Robotics} {Culturally sustainable social robotics}\ (\BPGS\ 632--642).
\newblock
\APACaddressPublisher{}{IOS Press}.
\PrintBackRefs{\CurrentBib}

\bibitem [\protect \citeauthoryear {%
Geyer%
}{%
Geyer%
}{%
{\protect \APACyear {2011}}%
}]{%
geyer2011introduction}
\APACinsertmetastar {%
geyer2011introduction}%
\begin{APACrefauthors}%
Geyer, C\BPBI J.%
\end{APACrefauthors}%
\unskip\
\newblock
\APACrefYearMonthDay{2011}{}{}.
\newblock
{\BBOQ}\APACrefatitle {{Introduction} to {Markov} {Chain} {Monte} {Carlo}} {{Introduction} to {Markov} {Chain} {Monte} {Carlo}}.{\BBCQ}
\newblock
\APACjournalVolNumPages{Handbook of {Markov} {Chain} {Monte} {Carlo}}{20116022}{}{45}.
\PrintBackRefs{\CurrentBib}

\bibitem [\protect \citeauthoryear {%
G{\"u}rb{\"u}zbalaban%
, Ozdaglar%
, Vanli%
\BCBL {}\ \BBA {} Wright%
}{%
G{\"u}rb{\"u}zbalaban%
\ \protect \BOthers {.}}{%
{\protect \APACyear {2020}}%
}]{%
gurbuzbalaban2020randomness}
\APACinsertmetastar {%
gurbuzbalaban2020randomness}%
\begin{APACrefauthors}%
G{\"u}rb{\"u}zbalaban, M.%
, Ozdaglar, A.%
, Vanli, N\BPBI D.%
\BCBL {}\ \BBA {} Wright, S\BPBI J.%
\end{APACrefauthors}%
\unskip\
\newblock
\APACrefYearMonthDay{2020}{}{}.
\newblock
{\BBOQ}\APACrefatitle {{Randomness} and {Permutations} in {Coordinate} {Descent} {Methods}} {{Randomness} and {Permutations} in {Coordinate} {Descent} {Methods}}.{\BBCQ}
\newblock
\APACjournalVolNumPages{Mathematical Programming}{181}{}{349--376}.
\PrintBackRefs{\CurrentBib}

\bibitem [\protect \citeauthoryear {%
Hecker%
}{%
Hecker%
}{%
{\protect \APACyear {2011}}%
}]{%
hecker2011my}
\APACinsertmetastar {%
hecker2011my}%
\begin{APACrefauthors}%
Hecker, C.%
\end{APACrefauthors}%
\unskip\
\newblock
\APACrefYearMonthDay{2011}{}{}.
\newblock
{\BBOQ}\APACrefatitle {{My} {Liner} {Notes} for {Spore}} {{My} {Liner} {Notes} for {Spore}}.{\BBCQ}
\newblock
\APACjournalVolNumPages{Retrieved September}{20}{}{2011}.
\PrintBackRefs{\CurrentBib}

\bibitem [\protect \citeauthoryear {%
Jozefowicz%
, Vinyals%
, Schuster%
, Shazeer%
\BCBL {}\ \BBA {} Wu%
}{%
Jozefowicz%
\ \protect \BOthers {.}}{%
{\protect \APACyear {2016}}%
}]{%
jozefowicz2016exploring}
\APACinsertmetastar {%
jozefowicz2016exploring}%
\begin{APACrefauthors}%
Jozefowicz, R.%
, Vinyals, O.%
, Schuster, M.%
, Shazeer, N.%
\BCBL {}\ \BBA {} Wu, Y.%
\end{APACrefauthors}%
\unskip\
\newblock
\APACrefYearMonthDay{2016}{}{}.
\newblock
{\BBOQ}\APACrefatitle {{Exploring} the {Limits} of {Language} {Modeling}} {{Exploring} the {Limits} of {Language} {Modeling}}.{\BBCQ}
\newblock
\APACjournalVolNumPages{arXiv preprint arXiv:1602.02410}{}{}{}.
\PrintBackRefs{\CurrentBib}

\bibitem [\protect \citeauthoryear {%
Kim%
\ \BBA {} Ruip{\'e}rez-Valiente%
}{%
Kim%
\ \BBA {} Ruip{\'e}rez-Valiente%
}{%
{\protect \APACyear {2020}}%
}]{%
kim2020data}
\APACinsertmetastar {%
kim2020data}%
\begin{APACrefauthors}%
Kim, Y\BPBI J.%
\BCBT {}\ \BBA {} Ruip{\'e}rez-Valiente, J\BPBI A.%
\end{APACrefauthors}%
\unskip\
\newblock
\APACrefYearMonthDay{2020}{}{}.
\newblock
{\BBOQ}\APACrefatitle {{Data-driven} {Game} {Design}: {The} {Case} of {Difficulty} in {Educational} {Games}} {{Data-driven} {Game} {Design}: {The} {Case} of {Difficulty} in {Educational} {Games}}.{\BBCQ}
\newblock
\BIn{} \APACrefbtitle {Addressing {Global} {Challenges} and {Quality} {Education}: 15th {European} {Conference} on {Technology} {Enhanced} {Learning}, {EC-TEL} 2020, {Heidelberg}, {Germany}, {September} 14--18, 2020, {Proceedings} 15} {Addressing {Global} {Challenges} and {Quality} {Education}: 15th {European} {Conference} on {Technology} {Enhanced} {Learning}, {EC-TEL} 2020, {Heidelberg}, {Germany}, {September} 14--18, 2020, {Proceedings} 15}\ (\BPGS\ 449--454).
\PrintBackRefs{\CurrentBib}

\bibitem [\protect \citeauthoryear {%
Laine%
\ \BBA {} Lindberg%
}{%
Laine%
\ \BBA {} Lindberg%
}{%
{\protect \APACyear {2020}}%
}]{%
laine2020designing}
\APACinsertmetastar {%
laine2020designing}%
\begin{APACrefauthors}%
Laine, T\BPBI H.%
\BCBT {}\ \BBA {} Lindberg, R\BPBI S.%
\end{APACrefauthors}%
\unskip\
\newblock
\APACrefYearMonthDay{2020}{}{}.
\newblock
{\BBOQ}\APACrefatitle {{Designing} {Engaging} {Games} for {Education}: A {Systematic} {Literature} {Review} on {Game} {Motivators} and {Design} {Principles}} {{Designing} {Engaging} {Games} for {Education}: A {Systematic} {Literature} {Review} on {Game} {Motivators} and {Design} {Principles}}.{\BBCQ}
\newblock
\APACjournalVolNumPages{IEEE Transactions on Learning Technologies}{13}{4}{804--821}.
\PrintBackRefs{\CurrentBib}

\bibitem [\protect \citeauthoryear {%
Leutner%
, Codreanu%
, Brink%
\BCBL {}\ \BBA {} Bitsakis%
}{%
Leutner%
\ \protect \BOthers {.}}{%
{\protect \APACyear {2023}}%
}]{%
leutner2023game}
\APACinsertmetastar {%
leutner2023game}%
\begin{APACrefauthors}%
Leutner, F.%
, Codreanu, S\BHBI C.%
, Brink, S.%
\BCBL {}\ \BBA {} Bitsakis, T.%
\end{APACrefauthors}%
\unskip\
\newblock
\APACrefYearMonthDay{2023}{}{}.
\newblock
{\BBOQ}\APACrefatitle {{Game} {Based} {Assessments} of {Cognitive} {Ability} in {Recruitment}: {Validity}, {Fairness} and {Test-taking} {Experience}} {{Game} {Based} {Assessments} of {Cognitive} {Ability} in {Recruitment}: {Validity}, {Fairness} and {Test-taking} {Experience}}.{\BBCQ}
\newblock
\APACjournalVolNumPages{Frontiers in {Psychology}}{13}{}{942662}.
\PrintBackRefs{\CurrentBib}

\bibitem [\protect \citeauthoryear {%
Liao%
\ \protect \BOthers {.}}{%
Liao%
\ \protect \BOthers {.}}{%
{\protect \APACyear {2022}}%
}]{%
liao2022fast}
\APACinsertmetastar {%
liao2022fast}%
\begin{APACrefauthors}%
Liao, Q.%
, Chen, J.%
, Wang, Z.%
, Bai, B.%
, Jin, S.%
\BCBL {}\ \BBA {} Wu, H.%
\end{APACrefauthors}%
\unskip\
\newblock
\APACrefYearMonthDay{2022}{}{}.
\newblock
{\BBOQ}\APACrefatitle {{F}ast {S}inkhorn {I}: {A}n {O} ({N}) algorithm for the {W}asserstein-1 metric} {{F}ast {S}inkhorn {I}: {A}n {O} ({N}) algorithm for the {W}asserstein-1 metric}.{\BBCQ}
\newblock
\APACjournalVolNumPages{arXiv preprint arXiv:2202.10042}{}{}{}.
\PrintBackRefs{\CurrentBib}

\bibitem [\protect \citeauthoryear {%
Majuri%
, Koivisto%
\BCBL {}\ \BBA {} Hamari%
}{%
Majuri%
\ \protect \BOthers {.}}{%
{\protect \APACyear {2018}}%
}]{%
majuri2018gamification}
\APACinsertmetastar {%
majuri2018gamification}%
\begin{APACrefauthors}%
Majuri, J.%
, Koivisto, J.%
\BCBL {}\ \BBA {} Hamari, J.%
\end{APACrefauthors}%
\unskip\
\newblock
\APACrefYearMonthDay{2018}{}{}.
\newblock
{\BBOQ}\APACrefatitle {{Gamification} of {Education} and {Learning}: A {Review} of {Empirical} {Literature}} {{Gamification} of {Education} and {Learning}: A {Review} of {Empirical} {Literature}}.{\BBCQ}
\newblock
\BIn{} \APACrefbtitle {Proceedings of the 2nd international GamiFIN conference, GamiFIN 2018.} {Proceedings of the 2nd international gamifin conference, gamifin 2018.}
\PrintBackRefs{\CurrentBib}

\bibitem [\protect \citeauthoryear {%
Malone%
}{%
Malone%
}{%
{\protect \APACyear {1981}}%
}]{%
malone1981toward}
\APACinsertmetastar {%
malone1981toward}%
\begin{APACrefauthors}%
Malone, T\BPBI W.%
\end{APACrefauthors}%
\unskip\
\newblock
\APACrefYearMonthDay{1981}{}{}.
\newblock
{\BBOQ}\APACrefatitle {{Toward} a {Theory} of {Intrinsically} {Motivating} {Instruction}} {{Toward} a {Theory} of {Intrinsically} {Motivating} {Instruction}}.{\BBCQ}
\newblock
\APACjournalVolNumPages{Cognitive Science}{5}{4}{333--369}.
\PrintBackRefs{\CurrentBib}

\bibitem [\protect \citeauthoryear {%
Manzano-Le{\'o}n%
\ \protect \BOthers {.}}{%
Manzano-Le{\'o}n%
\ \protect \BOthers {.}}{%
{\protect \APACyear {2021}}%
}]{%
manzano2021between}
\APACinsertmetastar {%
manzano2021between}%
\begin{APACrefauthors}%
Manzano-Le{\'o}n, A.%
, Camacho-Lazarraga, P.%
, Guerrero, M\BPBI A.%
, Guerrero-Puerta, L.%
, Aguilar-Parra, J\BPBI M.%
, Trigueros, R.%
\BCBL {}\ \BBA {} Alias, A.%
\end{APACrefauthors}%
\unskip\
\newblock
\APACrefYearMonthDay{2021}{}{}.
\newblock
{\BBOQ}\APACrefatitle {{Between} {Level} {Up} and {Game} {Over}: A {Systematic} {Literature} {Review} of {Gamification} in {Education}} {{Between} {Level} {Up} and {Game} {Over}: A {Systematic} {Literature} {Review} of {Gamification} in {Education}}.{\BBCQ}
\newblock
\APACjournalVolNumPages{Sustainability}{13}{4}{2247}.
\PrintBackRefs{\CurrentBib}

\bibitem [\protect \citeauthoryear {%
Match%
}{%
Match%
}{%
{\protect \APACyear {2022}}%
}]{%
Wordle}
\APACinsertmetastar {%
Wordle}%
\begin{APACrefauthors}%
Match, S.%
\end{APACrefauthors}%
\unskip\
\newblock
\APACrefYearMonthDay{2022}{}{}.
\newblock
{\BBOQ}\APACrefatitle {The {New} {York} {Times} {Buys} {Wordle}} {The {New} {York} {Times} {Buys} {Wordle}}.{\BBCQ}
\newblock
\APACjournalVolNumPages{The New York Times}{}{}{}.
\PrintBackRefs{\CurrentBib}

\bibitem [\protect \citeauthoryear {%
Park%
\ \protect \BOthers {.}}{%
Park%
\ \protect \BOthers {.}}{%
{\protect \APACyear {2023}}%
}]{%
park2023generative}
\APACinsertmetastar {%
park2023generative}%
\begin{APACrefauthors}%
Park, J\BPBI S.%
, O'Brien, J.%
, Cai, C\BPBI J.%
, Morris, M\BPBI R.%
, Liang, P.%
\BCBL {}\ \BBA {} Bernstein, M\BPBI S.%
\end{APACrefauthors}%
\unskip\
\newblock
\APACrefYearMonthDay{2023}{}{}.
\newblock
{\BBOQ}\APACrefatitle {{Generative} {Agents}: {Interactive} {Simulacra} of {Human} {Behavior}} {{Generative} {Agents}: {Interactive} {Simulacra} of {Human} {Behavior}}.{\BBCQ}
\newblock
\BIn{} \APACrefbtitle {Proceedings of the 36th Annual ACM Symposium on User Interface Software and Technology} {Proceedings of the 36th annual acm symposium on user interface software and technology}\ (\BPGS\ 1--22).
\PrintBackRefs{\CurrentBib}

\bibitem [\protect \citeauthoryear {%
Ryan%
\ \BBA {} Deci%
}{%
Ryan%
\ \BBA {} Deci%
}{%
{\protect \APACyear {2000}}%
}]{%
ryan2000self}
\APACinsertmetastar {%
ryan2000self}%
\begin{APACrefauthors}%
Ryan, R\BPBI M.%
\BCBT {}\ \BBA {} Deci, E\BPBI L.%
\end{APACrefauthors}%
\unskip\
\newblock
\APACrefYearMonthDay{2000}{}{}.
\newblock
{\BBOQ}\APACrefatitle {{Self-Determination} {Theory} and the {Facilitation} of {Intrinsic} {Motivation}, {Social} {Development}, and {Well-Being}} {{Self-Determination} {Theory} and the {Facilitation} of {Intrinsic} {Motivation}, {Social} {Development}, and {Well-Being}}.{\BBCQ}
\newblock
\APACjournalVolNumPages{American Psychologist}{55}{1}{68}.
\PrintBackRefs{\CurrentBib}

\bibitem [\protect \citeauthoryear {%
Sailer%
\ \BBA {} Homner%
}{%
Sailer%
\ \BBA {} Homner%
}{%
{\protect \APACyear {2020}}%
}]{%
sailer2020gamification}
\APACinsertmetastar {%
sailer2020gamification}%
\begin{APACrefauthors}%
Sailer, M.%
\BCBT {}\ \BBA {} Homner, L.%
\end{APACrefauthors}%
\unskip\
\newblock
\APACrefYearMonthDay{2020}{}{}.
\newblock
{\BBOQ}\APACrefatitle {{The} {Gamification} of {Learning}: A {Meta}-Analysis} {{The} {Gamification} of {Learning}: A {Meta}-analysis}.{\BBCQ}
\newblock
\APACjournalVolNumPages{Educational Psychology Review}{32}{1}{77--112}.
\PrintBackRefs{\CurrentBib}

\bibitem [\protect \citeauthoryear {%
Shaheen%
}{%
Shaheen%
}{%
{\protect \APACyear {2021}}%
}]{%
shaheen2021applications}
\APACinsertmetastar {%
shaheen2021applications}%
\begin{APACrefauthors}%
Shaheen, M\BPBI Y.%
\end{APACrefauthors}%
\unskip\
\newblock
\APACrefYearMonthDay{2021}{}{}.
\newblock
{\BBOQ}\APACrefatitle {{Applications} of {Artificial} {Intelligence} ({AI}) in {Healthcare}: {A} {Review}} {{Applications} of {Artificial} {Intelligence} ({AI}) in {Healthcare}: {A} {Review}}.{\BBCQ}
\newblock
\APACjournalVolNumPages{ScienceOpen Preprints}{}{}{}.
\PrintBackRefs{\CurrentBib}

\bibitem [\protect \citeauthoryear {%
Shang%
}{%
Shang%
}{%
{\protect \APACyear {2014}}%
}]{%
shang2014survey}
\APACinsertmetastar {%
shang2014survey}%
\begin{APACrefauthors}%
Shang, H\BPBI L.%
\end{APACrefauthors}%
\unskip\
\newblock
\APACrefYearMonthDay{2014}{}{}.
\newblock
{\BBOQ}\APACrefatitle {{A} {Survey} of {Functional} {Principal} {Component} {Analysis}} {{A} {Survey} of {Functional} {Principal} {Component} {Analysis}}.{\BBCQ}
\newblock
\APACjournalVolNumPages{AStA Advances in Statistical Analysis}{98}{}{121--142}.
\PrintBackRefs{\CurrentBib}

\bibitem [\protect \citeauthoryear {%
Solovyev%
, Bochkarev%
\BCBL {}\ \BBA {} Akhtyamova%
}{%
Solovyev%
\ \protect \BOthers {.}}{%
{\protect \APACyear {2019}}%
}]{%
solovyev2019google}
\APACinsertmetastar {%
solovyev2019google}%
\begin{APACrefauthors}%
Solovyev, V\BPBI D.%
, Bochkarev, V\BPBI V.%
\BCBL {}\ \BBA {} Akhtyamova, S\BPBI S.%
\end{APACrefauthors}%
\unskip\
\newblock
\APACrefYearMonthDay{2019}{}{}.
\newblock
{\BBOQ}\APACrefatitle {{Google} {Books} {Ngram}: {Problems} of {Representativeness} and {Data} {Reliability}} {{Google} {Books} {Ngram}: {Problems} of {Representativeness} and {Data} {Reliability}}.{\BBCQ}
\newblock
\BIn{} \APACrefbtitle {International {Conference} on {Data} {Analytics} and {Management} in {Data} {Intensive} {Domains}} {International {Conference} on {Data} {Analytics} and {Management} in {Data} {Intensive} {Domains}}\ (\BPGS\ 147--162).
\PrintBackRefs{\CurrentBib}

\bibitem [\protect \citeauthoryear {%
Sommerville%
}{%
Sommerville%
}{%
{\protect \APACyear {2020}}%
}]{%
sommerville2020social}
\APACinsertmetastar {%
sommerville2020social}%
\begin{APACrefauthors}%
Sommerville, J\BPBI A.%
\end{APACrefauthors}%
\unskip\
\newblock
\APACrefYearMonthDay{2020}{}{}.
\newblock
{\BBOQ}\APACrefatitle {{Social} {Cognition}} {{Social} {Cognition}}.{\BBCQ}
\newblock

\PrintBackRefs{\CurrentBib}

\bibitem [\protect \citeauthoryear {%
Teixeira%
, Wedel%
\BCBL {}\ \BBA {} Pieters%
}{%
Teixeira%
\ \protect \BOthers {.}}{%
{\protect \APACyear {2012}}%
}]{%
teixeira2012emotion}
\APACinsertmetastar {%
teixeira2012emotion}%
\begin{APACrefauthors}%
Teixeira, T.%
, Wedel, M.%
\BCBL {}\ \BBA {} Pieters, R.%
\end{APACrefauthors}%
\unskip\
\newblock
\APACrefYearMonthDay{2012}{}{}.
\newblock
{\BBOQ}\APACrefatitle {{Emotion-induced} {Engagement} in {Internet} {Video} {Advertisements}} {{Emotion-induced} {Engagement} in {Internet} {Video} {Advertisements}}.{\BBCQ}
\newblock
\APACjournalVolNumPages{Journal of Marketing Research}{49}{2}{144--159}.
\PrintBackRefs{\CurrentBib}

\bibitem [\protect \citeauthoryear {%
Van~Hove%
\ \protect \BOthers {.}}{%
Van~Hove%
\ \protect \BOthers {.}}{%
{\protect \APACyear {2019}}%
}]{%
van2019use}
\APACinsertmetastar {%
van2019use}%
\begin{APACrefauthors}%
Van~Hove, O.%
, Van~Muylem, A.%
, Leduc, D.%
, Legrand, A.%
, Jansen, B.%
, Feipel, V.%
\BDBL {}Bonnechere, B.%
\end{APACrefauthors}%
\unskip\
\newblock
\APACrefYearMonthDay{2019}{}{}.
\newblock
{\BBOQ}\APACrefatitle {The {Use} of {Cognitive} {Mobile} {Games} to {Assess} {Cognitive} {Function} of {Healthy} {Subjects} {Under} {Various} {Inspiratory} {Loads}} {The {Use} of {Cognitive} {Mobile} {Games} to {Assess} {Cognitive} {Function} of {Healthy} {Subjects} {Under} {Various} {Inspiratory} {Loads}}.{\BBCQ}
\newblock
\APACjournalVolNumPages{Medicine in {Novel} {Technology} and {Devices}}{1}{}{100005}.
\PrintBackRefs{\CurrentBib}

\bibitem [\protect \citeauthoryear {%
Wiley%
, Robinson%
, Mandryk%
\BCBL {}\ \protect \BOthers {.}}{%
Wiley%
\ \protect \BOthers {.}}{%
{\protect \APACyear {2021}}%
}]{%
wiley2021making}
\APACinsertmetastar {%
wiley2021making}%
\begin{APACrefauthors}%
Wiley, K.%
, Robinson, R.%
, Mandryk, R\BPBI L.%
\BCBL {}\ \BOthersPeriod {.}\end{APACrefauthors}%
\unskip\
\newblock
\APACrefYearMonthDay{2021}{}{}.
\newblock
{\BBOQ}\APACrefatitle {The {Making} and {Evaluation} of {Digital} {Games} {Used} for {The} {Assessment} of {Attention}: {Systematic} {Review}} {The {Making} and {Evaluation} of {Digital} {Games} {Used} for {The} {Assessment} of {Attention}: {Systematic} {Review}}.{\BBCQ}
\newblock
\APACjournalVolNumPages{JMIR {Serious} {Games}}{9}{3}{e26449}.
\PrintBackRefs{\CurrentBib}

\bibitem [\protect \citeauthoryear {%
Yen%
, Konold%
\BCBL {}\ \BBA {} McDermott%
}{%
Yen%
\ \protect \BOthers {.}}{%
{\protect \APACyear {2004}}%
}]{%
yen2004does}
\APACinsertmetastar {%
yen2004does}%
\begin{APACrefauthors}%
Yen, C\BHBI J.%
, Konold, T\BPBI R.%
\BCBL {}\ \BBA {} McDermott, P\BPBI A.%
\end{APACrefauthors}%
\unskip\
\newblock
\APACrefYearMonthDay{2004}{}{}.
\newblock
{\BBOQ}\APACrefatitle {{Does} {Learning} {Behavior} {Augment} {Cognitive} {Ability} as an {Indicator} of {Academic} {Achievement}?} {{Does} {Learning} {Behavior} {Augment} {Cognitive} {Ability} as an {Indicator} of {Academic} {Achievement}?}{\BBCQ}
\newblock
\APACjournalVolNumPages{Journal of {School} {Psychology}}{42}{2}{157--169}.
\PrintBackRefs{\CurrentBib}

\bibitem [\protect \citeauthoryear {%
Yin%
\ \protect \BOthers {.}}{%
Yin%
\ \protect \BOthers {.}}{%
{\protect \APACyear {2023}}%
}]{%
yin2023ai}
\APACinsertmetastar {%
yin2023ai}%
\begin{APACrefauthors}%
Yin, Q\BHBI Y.%
, Yang, J.%
, Huang, K\BHBI Q.%
, Zhao, M\BHBI J.%
, Ni, W\BHBI C.%
, Liang, B.%
\BDBL {}Wang, L.%
\end{APACrefauthors}%
\unskip\
\newblock
\APACrefYearMonthDay{2023}{}{}.
\newblock
{\BBOQ}\APACrefatitle {{AI} in {Human-Computer} {Gaming}: {Techniques}, {Challenges} and {Opportunities}} {{AI} in {Human-Computer} {Gaming}: {Techniques}, {Challenges} and {Opportunities}}.{\BBCQ}
\newblock
\APACjournalVolNumPages{Machine Intelligence Research}{20}{3}{299--317}.
\PrintBackRefs{\CurrentBib}

\bibitem [\protect \citeauthoryear {%
Zhao%
}{%
Zhao%
}{%
{\protect \APACyear {2017}}%
}]{%
zhao2017research}
\APACinsertmetastar {%
zhao2017research}%
\begin{APACrefauthors}%
Zhao, W.%
\end{APACrefauthors}%
\unskip\
\newblock
\APACrefYearMonthDay{2017}{}{}.
\newblock
{\BBOQ}\APACrefatitle {{Research} on {The} {Deep} {Learning} of {The} {Small} {Sample} {Data} {Based} on {Transfer} {Learning}} {{Research} on {The} {Deep} {Learning} of {The} {Small} {Sample} {Data} {Based} on {Transfer} {Learning}}.{\BBCQ}
\newblock
\BIn{} \APACrefbtitle {AIP {Conference} {Proceedings}} {Aip {Conference} {Proceedings}}\ (\BVOL\ 1864).
\PrintBackRefs{\CurrentBib}

\end{thebibliography}

\end{document}